# A CHARACTERIZATION OF THE DEGREE SEQUENCES OF 2-TREES

Prosenjit Bose[†]    Vida Dujmović[†]    Danny Krizanc[‡]    Stefan Langerman [§]    Pat Morin[†]
David R. Wood [¶]    Stefanie Wuhrer[†]

ABSTRACT. A graph $G$ is a 2-tree if $G = K_3$, or $G$ has a vertex $v$ of degree 2, whose neighbours are adjacent, and $G \setminus v$ is a 2-tree. A characterization of the degree sequences of 2-trees is given. This characterization yields a linear-time algorithm for recognizing and realizing degree sequences of 2-trees.

[†]School of Computer Science, Carleton University, Ottawa, Canada. Email:{jit,vida,morin,swuhrer}@scs.carleton.ca. The authors are partly supported by NSERC.

[‡]Department of Mathematics and Computer Science, Wesleyan University, Middletown,Connecticut, USA. Email:dkrizanc@caucus.cs.wesleyan.edu.

[§]Chercheur Qualifié du FNRS, Département d'Informatique, Université Libre de Bruxelles, Brussels, Belgium. Email:stefan.langerman@ulb.ac.be.

[¶]Departament de Matemàtica Aplicada II, Universitat Politècnica de Catalunya, Barcelona, Spain. Email:david.wood@upc.edu. The author is supported by a Marie Curie Fellowship of the European Community under contract 023865, and by the projects MCYT-FEDER BFM2003-00368 and Gen. Cat 2001SGR00224.

# 1 Introduction

The *degree sequence* of a graph[1] is the sequence[2] of the degrees of its vertices. If $D$ is the degree sequence of a graph $G$ then $G$ is a *realization* of $D$ and $G$ *realizes* $D$. Determining when a sequence of positive integers is realizable as a degree sequence of a simple graph has received much attention. The earliest result, by Erdős and Gallai [10], characterizes degree sequences of graphs. A non-increasing sequence of positive integers $d_1, \ldots, d_n$ is realizable if and only if $\sum_{i=1}^n d_i$ is even and $\sum_{i=1}^p d_i \leq p(p-1) + \sum_{i=p+1}^n \min(p, d_i)$ for all $p \leq n$. Hakimi [12] and Havel [15] give a strengthening of the result. In particular, they show that if a sequence is realizable then it is realizable by a graph in which a vertex of maximum degree is adjacent only to vertices of the highest degrees among the remaining vertices. Another generalization is derived by Cai *et al.* [4].

Degree sequences have been studied in connection with, among others, generating random graphs, extremal graph theory, and graph decompositions. Much work has gone into characterizing degree sequences of particular classes of graphs. That a sequence of $n$ positive integers is the degree sequence of a tree if and only if it sums to $2n - 2$ is a folklore result. Other graphs families with known degree sequence characterizations include split graphs [14, 20], $C_4$-minor[3] free graphs [21], unicyclic graphs [1], cacti graphs [15], Halin graphs [2], and edge-maximal outerplanar graphs [18]. The most investigated class of graphs is that of planar graphs. Despite the effort, no characterization of the degree sequences of planar graphs is known, even for edge-maximal planar graphs. Partial results are obtained in [5, 11, 13, 16].

A graph $G$ is a *k-tree* if either $G$ is the complete graph on $k + 1$ vertices, or $G$ has a vertex $v$ whose neighbourhood is a clique of order $k$ and the graph obtained by removing $v$ from $G$ is a $k$-tree. For example, 1-trees are trees.

In this paper we study the degree sequences of 2-trees. 2-trees are planar, and are the edge-maximal graphs with no $K_4$-minor [3]. Also, all edge-maximal outerplanar graphs are 2-trees, but not all 2-trees are outerplanar (consider $K_{2,3}$ for example). $k$-trees are intrinsically related to treewidth, which is an important parameter in the Robertson/Seymour theory of graph minors and in algorithmic complexity; see the surveys [3, 22]. In particular, a graph has a *treewidth* $k$ if and only if it is a subgraph of a $k$-tree. Thus $k$-trees are the edge-maximal graphs of treewidth $k$.

The following theorem is the main result of this paper. Let $a^{\langle b \rangle}$ denote the sequence $\langle a, \ldots, a \rangle$ of length $b$. A sequence of positive integers is *even* if all its elements are even.

**Theorem 1.** *Let $D$ be a sequence of $n$ integers. Let $n_2$ be the multiplicity of $2$ in $D$. Then $D$ is the degree sequence of a $2$-tree if and only if the following conditions are satisfied:*

(a) $\sum D = 4n - 6$,

(b) $\max D \leq n - 1$,

(c) $\min D = 2$ and $n_2 \geq 2$,

(d) $D \notin \{\langle 2^{\langle n-4 \rangle}, d^{\langle 4 \rangle} \rangle : d \geq 5\}$, *and*

---

[1] We consider graphs that are simple, finite, and undirected. The vertex set of a graph $G$ is denoted by $V(G)$, and its edge set by $E(G)$. The subgraph of $G$ induced by a set of vertices $S \subseteq V(G)$ is denoted by $G[S]$. $G \setminus S$ denotes $G[V(G) \setminus S]$.

[2] In this paper the term *sequence* will be used in place of multiset.

[3] A graph $H$ is a *minor* of a graph $G$ if $H$ is isomorphic to a graph obtained from a subgraph of $G$ by contracting edges.

(e) $n_2 \geq \frac{n}{3} + 1$ whenever $D$ is even.

*Moreover, if $D$ satisfies Conditions (a)–(e) then given any $\ell \in D$ such that $\ell \geq 3$, there exists a 2-tree that realizes $D$ in which a vertex of degree $\ell$ is adjacent to a vertex of degree 2.*

We denote by $\boldsymbol{\Delta}$ the set of all sequences satisfying Conditions (a)–(e) of Theorem 1 (hereafter simply referred to as Conditions (a)–(e)).

Independently, Lotker *et al.* [19] also studied degree sequences of 2-trees. Their main result is that if a sequence $D$ contains a 3, then Conditions (a)-(c) are sufficient for $D$ to be realizable as a 2-tree. This result is an immediate corollary of Theorem 1. By counting the number of sequences that satisfy (a)-(c) and contain a 3, they show that nearly every sequence that satisfies Conditions (a)-(c) is the degree sequence of some 2-tree.

A discussion on why it may be difficult to generalize our results to $k$-trees for general $k$, can be found in Section 6, along with more relevant results. In Section 2 we consider degree sequences of trees. In Section 3 we show that the degree sequence of every 2-tree is in $\boldsymbol{\Delta}$. In Section 4 we show that every sequence in $\boldsymbol{\Delta}$ is the degree sequence of a 2-tree. Section 5 discusses a linear-time algorithm for recognizing and realizing degree sequences of 2-trees.

## 2 Degree Sequences of Trees

The following lemma is a strengthening of the folklore characterization of the degree sequences of trees. We make use of this strengthening in Lemma 13.

**Lemma 1.** *Let $D$ be a sequence of $n$ positive integers. Then $D$ is the degree sequence of a tree if and only if $\sum D = 2n - 2$. Moreover, if $\sum D = 2n - 2$, then for any $\ell, k \in D$, $D$ can be realized as a tree in which a vertex of degree $\ell$ is adjacent to a vertex of degree $k$, unless $n > 2$ and $\ell = k = 1$.*

*Proof.* Every tree on $n$ vertices has $n - 1$ edges; thus its degrees sum to $2n - 2$.

Assume now that $D$ is a sequence of positive integers that sum to $2n - 2$. Assume $n \geq 3$, since for $n \leq 2$ the statement of the lemma is trivial.

We first prove by induction that $D$ is the degree sequence of a tree. For the induction step, $n \geq 3$, notice that since $\sum D = 2n - 2$ and since $n \geq 3$, there is at least one 1 in $D$ and at least one number, $x$, greater than 1 in $D$. Create a new sequence $D'$ from $D$ by removing one 1 and reducing $x$ by 1. $D'$ is comprised of $n - 1 \geq 2$ positive integers that sum to $2(n - 1) - 2$. By the inductive hypothesis, there exists a tree $T'$ that realizes $D'$. Adding a vertex to $T'$ adjacent to a vertex of degree $x - 1$, creates a tree $T$ on $n$ vertices that realizes $D$.

Now we prove the stronger claim. Assume without loss of generality that $\ell \leq k$. Let $T$ be a tree that realizes $D$. Let $y$ be a vertex of degree $\ell$ and $r$ a vertex of degree $k$ in $T$. If $ry$ is an edge in $T$ we are done. Otherwise, root $T$ at $r$. Since $n > 2$, $k \geq 2$, and thus $r$ has at least two children. Denote by $T_y$ the subtree of $T$ rooted at $y$ and by $p$ the parent of $y$ in $T$. Denote by $T_x$ a subtree of $T$ rooted at a child $x$ of $r$ that does not contain $y$; that is, $y \notin V(T_x)$. Now swap $T_x$ and $T_y$, as illustrated in Figure 1. In particular, delete edge $rx$, delete edge $py$, add edge $px$ and finally add edge $ry$. The resulting graph is a tree that realizes $D$ and has a vertex of degree $\ell$ adjacent to a vertex of degree $k$. □

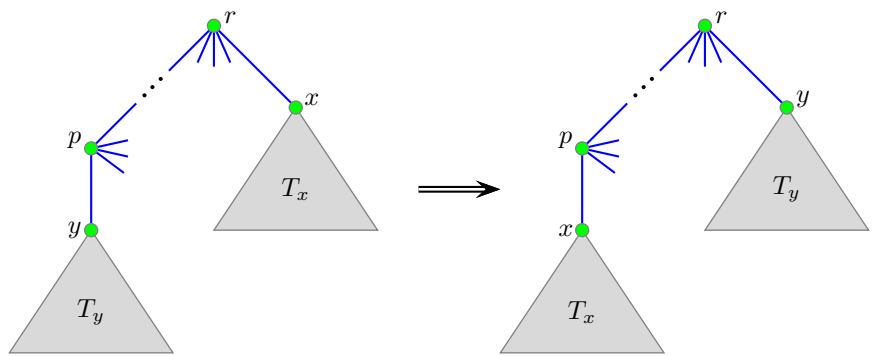

Figure 1: Illustration for the proof of Lemma 1.

## 3 Degree Sequences of 2-Trees are in $\Delta$

Several well-known properties of 2-trees are summarized in the following two lemmas, which can be proved by elementary inductive arguments.

An *ear* in a graph is a vertex of degree 2 whose neighbours are adjacent. A graph $G$ is a *2-tree* if $G = K_3$, or $G$ has an ear $u$ such that $G' := G \setminus u$ is a 2-tree. In other words, every 2-tree $G \neq K_3$ can be obtained from some 2-tree $G'$ by adding a new vertex $u$ adjacent to two vertices, $v$ and $w$, where $vw \in E(G')$. We call this process *attaching* the vertex $u$ to the edge $vw$.

**Lemma 2.** *Every 2-tree $G$ on $n$ vertices has the following properties:*

1. The sum of the degrees of the vertices in $G$ is $4n - 6$.
2. The minimum vertex degree of $G$ is 2.
3. Every vertex of degree 2 in $G$ is an ear.
4. $G$ has at least two ears.
5. No two ears in $G$ are adjacent unless $G = K_3$.
6. $G$ has no $K_4$-minor.
7. $G$ is 2-connected.

**Lemma 3.** *Let $T$ be a tree on at least two vertices. Then the graph $G$ obtained by adding a new vertex adjacent to each vertex of $T$ is a 2-tree.*

Lemma 2 along with Lemma 4 and Lemma 6 below prove that Conditions (a)-(e) are necessary in Theorem 1; that is, the degree sequence of every 2-tree is in $\Delta$.

**Lemma 4.** *For all $d \geq 5$, the sequence $D = \langle 2^{\langle n-4 \rangle}, d, d, d, d \rangle$ is not the degree sequence of a 2-tree.*

*Proof.* Suppose for the sake of contradiction that $G$ is a 2-tree that realizes $D$. Removing all ears from $G$ yields a 2-tree $G'$ on four vertices. The only 2-tree on four vertices is $K_4$ minus an edge, as depicted by the thick edges in Figure 2. Let $v_1, v_2, v_3, v_4$ be the vertices of $G'$, where $v_1 v_3$ is the only non-edge. Let $d_i(=d)$ be the degree of each vertex $v_i$ in $G$. Let $x_{i,j}$ be the number of ears attached to each edge $v_i v_j$. Then

$$d_1 + d_3 = (x_{1,4} + x_{1,2} + 2) + (x_{3,4} + x_{2,3} + 2) < (x_{1,2} + x_{2,4} + x_{2,3} + 3) + (x_{1,4} + x_{2,4} + x_{3,4} + 3) = d_2 + d_4,$$

which is not possible since $d_1 + d_3 = d_2 + d_4 = 2d$. $\square$

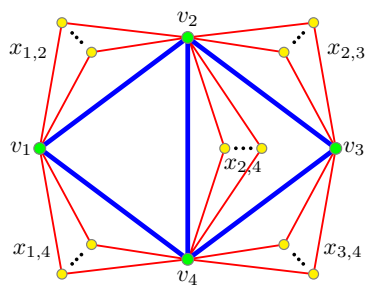

Figure 2: Illustration for the proof of Lemma 4.

To prove that Condition (e) is necessary we need the following lemma. Say an edge $vw$ is *close* to a vertex $u$ if both $v$ and $w$ are adjacent to $u$.

**Lemma 5.** *Let $G$ be a 2-tree with $n \geq 4$ vertices such that each edge is close to at most one ear. Then $G$ has an edge close to exactly two vertices, one of which is an ear.*

*Proof.* Let $S$ be the set of ears in $G$. Since $n \geq 4$, no two vertices in $S$ are adjacent in $G$ and $|S| \geq 2$, by Lemma 2. Consider the graph $G' := G \setminus S$. If $G' = K_2$, then the edge of $G'$ is close to at least two ears in $G$, which contradicts our assumption. Therefore, $G'$ has at least three vertices and $G'$ is a 2-tree. No pair of vertices in $S$ attaches to the same edge of $G'$ in $G$, as again that would contradict our assumption. Since $G'$ is a 2-tree, it has an ear, $v$. Since $v$ has degree greater than 2 in $G$, there is an edge $vw \in E(G')$ such that exactly one vertex, $u$, of $S$ attaches to $vw$ in $G$. Thus $vw$ is close to $u$ in $G$. Since $v$ is an ear in $G'$, $vw$ is close to exactly one vertex $y$ in $G'$. Since every vertex in $G'$ has degree greater than 2 in $G$, $y$ has degree at least three. This completes the proof, since $vw$ is close to exactly $u$ and $y$ in $G$. $\square$

**Lemma 6.** *Let $G$ be a 2-tree with $n$ vertices, of which $n_2$ are ears. If each vertex in $G$ has even degree then $n_2 \geq \frac{n}{3} + 1$.*

*Proof.* We proceed by induction on $n$. The base case with $n \leq 4$ is trivial. Now assume that $n \geq 5$.

Suppose that $G$ has an edge $vw$ close to at least two ears, $x$ and $y$. Let $G' := G \setminus \{x,y\}$. Since $n \geq 5$, $G'$ is a 2-tree each vertex of which has even degree. Say $G'$ has $n'$ vertices, of which $n'_2$ are ears. By induction $n'_2 \geq \frac{n'}{3} + 1$. Now $n_2 \geq n'_2 + 1$, since going from $G'$ to $G$, we attach two ears, $x$ and $y$, and delete at most one (since $v$ and $w$ cannot both be ears in $G'$, unless $G' = K_3$ in which case the result is immediate). Thus $n_2 \geq \frac{n'}{3} + 1 + 1 > \frac{n}{3} + 1$.

Now assume that each edge is close to at most one ear. By Lemma 5, $G$ has an edge $vw$ close to exactly two vertices $x$ and $y$, one of which, say $x$, is an ear. Thus $G[\{v,w,y\}] = K_3$. Consider the components of $G \setminus \{v,w,y\}$. For each component $C$, exactly two vertices in $\{v,w,y\}$ have a neighbour in $C$ (otherwise $G$ has a $K_4$-minor or a cut vertex which is impossible by Lemma 2). We say $C$ *attaches* to the edge between that pair of vertices. The only component that attaches to $vw$ is $x$ (otherwise $vw$ is close to more than two vertices). These concepts are illustrated in Figure 3.

Let $G'$ be the subgraph of $G$ induced by $v, y$ and the components of $G \setminus \{v,w,y\}$ that attach to $vy$. Let $G''$ be the subgraph of $G$ induced by $w, y$ and the components that attach to $wy$, as illustrated in Figure 3. Then degree of $v$ is even in $G'$ since it differs by two from its degree in $G$. Thus the degree of $y$ in $G'$ is even (otherwise $G'$ has exactly one vertex with odd degree which is impossible). Hence all vertices in $G'$ have even degrees. The same is true for $G''$. Say $G'$ has $n'$ vertices, of which $n'_2$ are ears, and $G''$ has $n''$ vertices, of which $n''_2$ are ears.

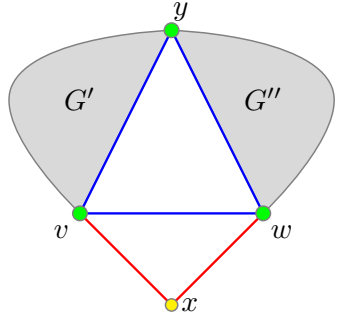

Figure 3: Illustration for the proof of Lemma 6.

Let $t'$ and $t''$, respectively, be the number of ears in $G'$ and $G''$ that have degree at least 3 in $G$. Hence

$$n_2 \geq (n'_2 - t') + (n''_2 - t'') + 1 \ , \tag{1}$$

where the "+1" is for $x$ which is neither in $G'$ nor $G''$. The only vertices with differing degrees in $G'$ and $G$ are $v$ and $y$. Since $v$ and $y$ are adjacent in $G'$, at most one of $v$ and $y$ has degree two in $G'$ (unless $G' = K_3$). That is, $t' \leq 1$ (unless $G' = K_3$ in which case $t' = 2$). By induction, $n'_2 \geq \frac{n'}{3} + 1$. Moreover, if $G' = K_3$ then $n'_2 = \frac{n'}{3} + 2$. Thus $n'_2 - t' \geq \frac{n'}{3}$. Similarly, $n''_2 - t'' \geq \frac{n''}{3}$. By Equation (1), $n_2 \geq \frac{n'}{3} + \frac{n''}{3} + 1 = \frac{n}{3} + 1$ as desired. □

The results of this section prove the following lemma.

**Lemma 7.** *The degree sequence of every 2-tree is in* **Δ**.

# 4 The Elements of $\Delta$ are the Degree Sequences of 2-Trees

In this section we prove that every $D \in \Delta$ is the degree sequence of a 2-tree. Our proof is by induction on $n$, the length of the sequence.

## 4.1 The Base Cases

In this section we give constructions for the base cases that occur in our inductive proof. The proofs ignore the number of ears. However, Condition (a) and Lemma 2.1 imply that the constructed 2-trees have the correct number of ears.

**Lemma 8.** *The sequence $\langle 2, 2, 2 \rangle$ is the degree sequences of a 2-tree $K_3$.*

**Lemma 9.** *Let $D$ be a sequence of $n$ integers such that $D \in \Delta \cap \{\langle 2^{\langle n-2 \rangle}, x, y \rangle : x, y \geq 3\}$. Then there exists a 2-tree that realizes $D$ in which every vertex of degree greater than 2 is adjacent to an ear.*

*Proof.* From Condition (a) we know that $2(n-2) + x + y = 4n - 6$ or, equivalently, $x + y = 2n - 2$. By Condition (b) this implies that $x = y = n - 1$. Thus, we can create a 2-tree realizing $D$ by starting from $K_3$ and attaching $n - 3$ vertices to one of its edges. Clearly, in the resulting 2-tree, every vertex of degree greater than 2 is adjacent to an ear. $\square$

**Lemma 10.** *Let $D$ be a sequence of $n$ integers such that $D \in \Delta \cap \{\langle 2^{\langle n-3 \rangle}, x, y, z \rangle : x, y, z \geq 3\}$. Then there exists a 2-tree that realizes $D$ in which every vertex of degree greater than 2 is adjacent to an ear.*

*Proof.* Create a 2-tree by attaching $e_i$ vertices to the $i$-th edge of $K_3$, where

$$e_1 = \tfrac{1}{2}(x+y-z-2) \ , \ \ e_2 = \tfrac{1}{2}(x-y+z-2) \ , \ \ e_3 = \tfrac{1}{2}(-x+y+z-2) \ .$$

It is straightforward to verify that the resulting 2-tree has three vertices of degree $x$, $y$, and $z$, respectively, and that all other vertices are ears.

It remains to verify that $e_1$, $e_2$ and $e_3$ are non-negative integers. These numbers are certainly integers because, by Condition (a), $2(n-3) + x + y + z = 4n - 6$ or, equivalently, $x + y + z = 2n$. Next we show that $e_1$ is non-negative. By Condition (b), $x + y \geq n + 1$ and $x + y - z \geq 2$. Thus $x + y - z - 2 \geq 0$ and $e_1$ is non-negative, as required. An analogous argument shows that $e_2$ and $e_3$ are also non-negative. $\square$

**Lemma 11.** *Let $D$ be a sequence of $n$ integers such that $D \in \Delta \cap \{\langle 2^{\langle n-5 \rangle}, x, d, d, d, d+x-2 \rangle : x \geq 3, d \geq 5\}$. Then there exists a 2-tree that realizes $D$ in which every vertex of degree greater than 2 is adjacent to an ear.*

*Proof.* Begin with the 2-tree on five vertices, as depicted by the thick edges in Figure 4(a), where one vertex, $v_1$, has degree 4, two vertices, $v_3$ and $v_4$, have degree 3 and two vertices, $v_2$ and $v_5$, are ears. Attach one vertex to $v_2v_3$ and one to $v_4v_5$, attach $d - 4 > 0$ vertices to $v_3v_4$, attach $x - 3 \geq 0$ vertices to $v_1v_2$ and attach $d - 3 > 0$ vertices to $v_1v_5$. Then the degree of $v_1$ is $d + x - 2$, the degree of $v_2$ is $x$, and the degrees of $v_3$, $v_4$ and $v_5$ are $d$. All other vertices are ears, as required. In addition, each vertex of degree greater than 2 is adjacent to an ear. $\square$

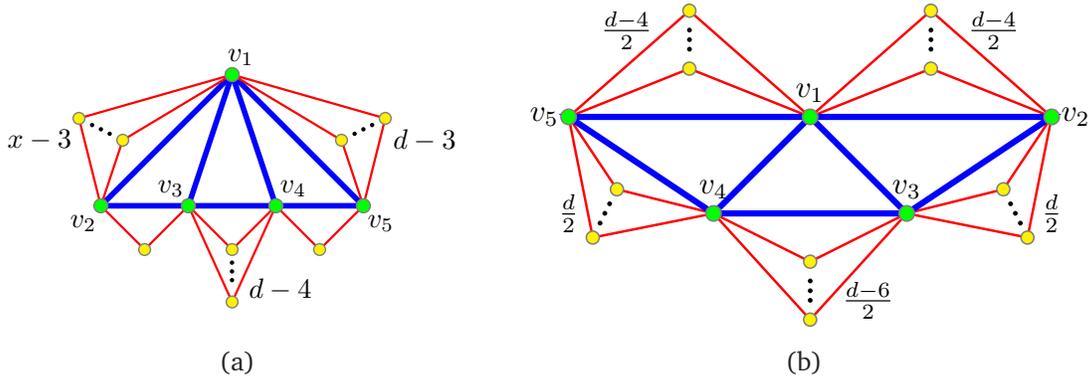

Figure 4: Illustration for the proof of (a) Lemma 11, and (b) Lemma 12.

**Lemma 12.** *Let $D$ be a sequence of $n$ integers such that $D \in \Delta \cap \{\langle 2^{\langle n-5 \rangle}, d^{\langle 5 \rangle}\rangle : d \geq 5\}$. Then there exists a 2-tree that realizes $D$ in which every vertex of degree $d$ is adjacent to an ear.*

*Proof.* Start with the 2-tree induced by the thick edges in Figure 4(b). By Condition (a), $2(n-5)+5d = 4n-6$; that is, $5d = 2(n+2)$. Thus $d$ is even. Attach $\frac{d-4}{2} \geq 1$ vertices to each of the edges $v_1v_2$ and $v_1v_5$. Attach $\frac{d-6}{2} \geq 0$ vertices to the edge $v_3v_4$, and $\frac{d}{2} > 0$ vertices to each of the edges $v_2v_3$ and $v_4v_5$. The resulting 2-tree $G$ has five vertices of degree $d$ and each vertex of degree $d$ is adjacent to an ear. Thus $G$ is a desired realization of $D$. □

**Lemma 13.** *Let $D$ be a sequence of $n$ integers with $\min D > 1$ and $\sum D = 4n - 6$. If $n - 1 \in D$, then for any $\ell, k \in D$, $D$ can be realized as a 2-tree in which a vertex of degree $\ell$ is adjacent to a vertex of degree $k$, unless $n > 3$ and $\ell = k = 2$.*

*Proof.* No sequence of integers greater than 1 sum to $4n-6$ if $n < 3$. If $n = 3$, only $D = \{2, 2, 2\}$ meets the criteria and the claimed lemma is correct by Lemma 8. Therefore we may assume that $n \geq 4$, $\ell \geq k$, and $\ell \geq 3$.

Let $D'$ be the sequence obtained from $D$ by removing $n-1$ from $D$ and reducing each remaining number by 1. $D'$ is comprised of $n' := n - 1 \geq 3$ positive integers that sum to $2n' - 2$.

First consider the case that $\ell = n - 1$. By Lemma 1, $D'$ can be realized by a tree. By Lemma 3, we build a 2-tree that realizes $D$ with one vertex of degree $\ell = n - 1$ adjacent to all other vertices.

Now consider the case that $\ell < n - 1$. Then $k - 1, \ell - 1 \in D'$. Since $\ell - 1 \geq 2$, $D'$ can be realized as a tree where a vertex of degree $k - 1$ is adjacent to a vertex of degree $\ell - 1$, as implied by Lemma 1. Again, add a vertex $v$ and an edge between $v$ and each vertex of the tree. By Lemma 3, the resulting graph $G$ is a 2-tree that realizes $D$ in which a vertex of degree $k$ is adjacent to a vertex of degree $\ell$. □

**Lemma 14.** *Let $D \in \Delta$ be a sequence of $n$ integers. Let $n_2$ be the multiplicity of 2 in $D$. For any $x, y \in D$, such that $x \geq 3$ and $x \neq y$, if there exists an integer $r$ such that $1 \leq r \leq n_2$, $x - r \geq 2$, $y - r \geq 2$ and* $n - r - 1 \in D$, then $D$ can be realized as a 2-tree in which an ear is adjacent to a vertex of degree $x$ and a vertex of degree $y$.

*Proof.* Let $D'$ be the sequence of length $n'$ obtained from $D$ by removing $r$ 2's from $D$ and by reducing both $x$ and $y$ by $r$. $D'$ is comprised of $n' = n - r$ integers greater than 1 that sum to $4n' - 6$ and $n' - 1 \in D'$. Since $x \neq y$, at least one of $x - r$ and $y - r$ is greater than 2. All this implies that Lemma 13 is applicable to $D'$ with $\ell = x - r$ and $k = y - r$. Therefore, $D'$ can be realized as a 2-tree $G'$ in which a vertex of degree $x - r$ is adjacent to a vertex of degree $y - r$. Attaching $r \geq 1$ ears to that edge, gives a 2-tree that realizes $D$ in which an ear is adjacent to a vertex of degree $x$ and a vertex of degree $y$. □

## 4.2 The Induction

With the base cases out of the way, we are ready for an inductive proof of the sufficiency of Conditions (a)-(e) in Theorem 1.

**Lemma 15.** *Suppose that $D \in \Delta$. For each $\ell \in D$ such that $\ell \geq 3$, there exists a 2-tree that realizes $D$ in which a vertex of degree $\ell$ is adjacent to an ear.*

*Proof.* Let $n$ denote the number of elements of $D$ and let $n_t$ denote the multiplicity of $t$ in $D$.

We are given $D$ and a particular value $\ell \in D$. If $D$ meets the conditions of Lemmas 8, 9, 10, 11, 12, 13, or 14 then we are done, and we say that $D$ is a base case. Otherwise, we proceed as follows. Below we select a value $k \in D$ such that $k \geq 3$. Then we create a new sequence $D' \in \Delta$ of length $n' < n$ to which we can apply induction. From a realization of $D'$, we construct a 2-tree that realizes $D$ in which a vertex of degree $\ell$ is adjacent to an ear and a vertex of degree $k$ is adjacent to an ear. The choice of $k$ and the reduction needed to obtain $D'$ depends on $D$ and $\ell$. We distinguish the following cases.

We say that $D$ is *flat* if $D$ has at most two distinct elements. $D$ is *special* if it is not flat and if one of the following is true:

(a) $D$ is even and $n_4 \geq 3$, or

(b) $D$ is not even, $n_4 \geq 3$, and $D$ has exactly two odd numbers one of which is 3 and the other is $x \geq 5$, and $\ell = x$.

Finally, $D$ is *typical* if it is neither flat nor special.

Before describing how we choose $k$ and perform a reduction to $D'$, we make the following observations.

**Observation 1.** *If $D \in \Delta$ is flat and not a base case, then $D = \langle 2^{\langle n_2 \rangle}, d^{\langle n_d \rangle} \rangle$ such that $d \geq 5$, $n_2 \geq 3d - 6$, $n_d \geq 6$ and $n \geq 3d$.*

*Proof.* Since $D$ has at most two distinct elements, and by Condition (c), one element is 2, $D = \langle 2^{\langle n_2 \rangle}, d^{\langle n_d \rangle} \rangle$, for some $d \geq 2$. If $d = 2$ then $D = \langle 2, 2, 2 \rangle$ by Condition (a), and $D$ is the base case in Lemma 8. Also $\langle 2, 2, 2 \rangle$ is the only sequence in $\boldsymbol{\Delta}$ with $n \leq 3$. Now assume that $d \geq 3$ and $n \geq 4$. Thus Condition (b) implies $n_d \geq 2$, as otherwise $2(n - 1) + d = 4n - 6$ which gives $d \geq n$ whenever $n \geq 4$. Since $D$ does not meet the conditions of Lemmas 9 and 10, $n_d \geq 4$. Then Condition (a) implies that $d \geq 4$. Suppose that $d = 4$. Condition (a) is equivalent to $n_d(d - 4) + 6 = 2n_2$ which implies that $n_2 = 3$. Since $n_d \geq 4$, that implies $n \geq 7$. Thus $D = \langle 2^{\langle 3 \rangle}, 4^{\langle n-3 \rangle} \rangle$ with $n \geq 7$. This sequence is excluded from $\boldsymbol{\Delta}$ by Condition (e). Thus we may assume $n_d \geq 4$ and $d \geq 5$. The case $n_d = 4$ and $d \geq 5$ is excluded by Condition (d). The case $n_d = 5$ and $d \geq 5$ is handled in Lemma 12. Thus $d \geq 5$ and $n_d \geq 6$ as required.

Condition (a) when applied to $D$ gives $dn_d + 2n_2 = 4(n_2 + n_d) - 6$, which simplifies to $n_2 = \frac{n_d}{2}(d - 4) + 6 \geq 3(d - 4) + 6 = 3d - 6$ and $n \geq 3d$. $\square$

Let $\alpha$ be the minimum integer in $D$ greater than 2.

**Observation 2.** *If $D \in \boldsymbol{\Delta}$ is not a base case, then $n_2 \geq \alpha - 1$.*

*Proof.* If $\alpha = 3$ then Condition (c) gives $n_2 \geq 2 = \alpha - 1$. Otherwise, (as in the proof of Observation 1) Condition (b) implies $n - n_2 \geq 2$ and thus there are at least two elements of $D$ greater or equal to $\alpha$. Therefore, Condition (a) gives $2n_2 + \alpha + \alpha + 4(n - n_2 - 2) \leq 4n - 6$ which simplifies to $n_2 \geq \frac{\alpha + \alpha}{2} - 1 \geq \alpha - 1$, as required. $\square$

The value $k$ is selected as follows. If $D$ is flat, $k := \alpha = d$. Otherwise, choose $k$ such that $k \neq \ell$. In particular, if $D$ is typical or $D$ is special and even, choose $k := \alpha$. If $\ell = \alpha$, then redefine $\ell$ to be the smallest number greater than $\alpha$ in $D$, thus reversing the roles of $\ell$ and $k$. (We are allowed to do this since the 2-tree realizing $D$ that we construct has a vertex of degree $\ell$ and a vertex of degree $k$ and each is adjacent to an ear). Otherwise, $D$ is special and not even and we choose $k := 4$. Thus unless $D$ is flat, $k < \ell$. Also note that whenever $D$ is special, $k = 4$.

We now create a new sequence $D'$ of length $n' < n$ to which we can apply induction.

- If $D$ is flat, then create $D'$ by removing $2d - 7$ 2's and two $d$'s from $D$, and reducing one $d$ by $d - 2$ and one $d$ by $d - 4$. By Observation 1, $n_d \geq 6$, $d - 4 > 0$ and $n_2 > 2d - 7$. Thus $D'$ is well defined and all of its elements are positive integers greater than 1.

- If $D$ is typical, then create $D'$ by removing $k - 2$ 2's from $D$, and reducing both $\ell$ and $k$ by $k - 2$. Observation 2, in particular having $n_2 \geq \alpha - 1$ and $k = \alpha$, implies $k - 2 < n_2$. Thus $D'$ is well defined and all of its elements are positive integers greater than 1.

- If $D$ is special, then create $D'$ by removing two 2's and one 4 from $D$ and reducing both $\ell$ and $k$ by 2. By the choice of $k$, $k = \alpha$ or $k = 4$. If $k = \alpha$, then Observation 2 implies $k - 2 \leq n_2$; and, if $k = 4$, then Condition (c) implies $k - 2 \leq n_2$. Thus $D'$ is well defined and all of its elements are positive integers greater than 1.

The proof of the following claim is left until later.

**Claim 1.** $D' \in \boldsymbol{\Delta}$.

If $D$ is flat, apply the inductive hypothesis to the sequence $D'$ with the special value 4 to obtain a 2-tree $G'$ in which a vertex $v$ of degree 4 is adjacent to an ear $w$. Attach $d - 4 > 0$ vertices to $vw$ and call one of them $u$. Attach one vertex, $q$, to the edge $wu$. Attach one vertex to the edge $wq$ and $d - 3 > 0$ vertices to the edge $uq$. This construction is illustrated in Figure 5(a). The resulting 2-tree $G$ is a realization of $D$ in which an ear is adjacent to vertices of degree $\ell = d$ and $k = d$ (consider for example $w$ and $q$).

Otherwise, if $D$ is not flat, apply the inductive hypothesis to the sequence $D'$ with the special value $\ell - k + 2 \geq 3$ to obtain a 2-tree $G'$ in which a vertex $v$ of degree $\ell - k + 2$ is adjacent to an ear $w$. If $D$ is typical, attach $k - 2 \geq 1$ vertices to the edge $vw$ to obtain a 2-tree $G$, as illustrated in Figure 5(b). Otherwise, $D$ is special, first attach one vertex $u$ to the edge $vw$. Then attach one vertex to the edge $vu$ and one to the edge $wu$, as illustrated in Figure 5(c). In both cases the resulting 2-tree $G$ is a realization of $D$ in which a vertex, $v$, of degree $\ell$ is adjacent to an ear and a vertex, $w$, of degree $k$ is adjacent to an ear. This completes the proof. □

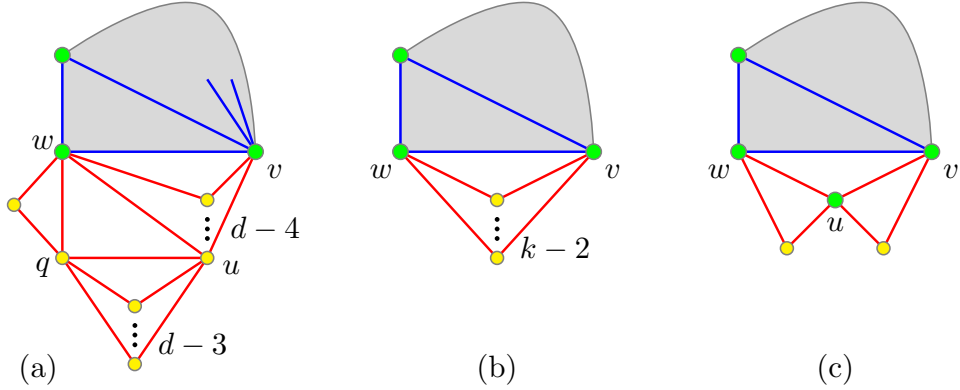

Figure 5: The induction step when (a) $D$ is flat, (b) $D$ is typical, (c) $D$ is special.

It remains to prove Claim 1.

*Proof of Claim 1.* By the construction of $D'$, it is clear that all elements in $D'$ are positive integers greater than 1. We must show that $D'$ satisfies Conditions (a)-(e). We know that $D$ satisfies Conditions (a)-(e) and that $D$ does not satisfy the conditions of any of Lemmas 8–14. Let $n'$ denote the number of elements of $D'$ and let $n'_t$ denote the multiplicity of $t$ in $D'$.

*Proof that $D'$ satisfies Condition (a); that is, $\sum D' = 4n' - 6$:*

If $D$ is flat, then $n' = n - 2d + 7 - 2 = n - 2d + 5$. Since $D$ satisfies Condition (a), $\sum D' = (4n - 6) - 2(2d - 7) - 2d - (d - 2) - (d - 4) = 4(n - 2d + 5) - 6 = 4n' - 6$.

If $D$ is typical, then $n' = n - k + 2$. Since $D$ satisfies Condition (a), $\sum D' = (4n - 6) - 2(k - 2) - (k - 2) - (k - 2) = 4(n - k + 2) - 6 = 4n' - 6$.

Otherwise, $D$ is special and $n' = n-3$ and $\sum D' = (4n-6) - 2 \cdot 2 - 4 - 2 \cdot 2 = 4(n-3) - 6 = 4n' - 6$.

*Proof that $D'$ satisfies Condition (c); that is, $\min D' = 2$ and $n'_2 \geq 2$:*

It is clear from the construction that $\min D' = 2$ in each case.

If $D$ is flat, $n'_2 = n_2 - (2d-7) + 1$. Thus Observation 1 implies $n'_2 \geq 2$. If $D$ is typical, then $\alpha = k$, and $n'_2 = n_2 - k + 2 + 1 = n_2 - \alpha + 3$. Then Observation 2, in particular, having $n_2 \geq \alpha - 1$ implies $n'_2 \geq \alpha - 1 - \alpha + 3 = 2$. If $D$ is special and even, then $\alpha = k = 4$, and $n'_2 = n_2 - 2 + 1 = n_2 - 1$. Since $n_2 \geq \alpha - 1$, $n'_2 \geq \alpha - 2 = 2$. Finally, consider the case that $D$ is special and not even. Then $\alpha = 3$, $k = 4$, and $n'_2 = n_2 - 2 + 1 = n_2 - 1$. Thus it is enough to prove that in this case $n_2 \geq 3$. From Condition (a) when applied to $D$ and since $D$ is special and not even, we have $2n_2 + 3 + 4(n - n_2 - 2) + \ell \leq 4n - 6$, which simplifies to $n_2 \geq \frac{\ell+1}{2}$. Since $\ell \geq 5$, $n_2 \geq 3$, as required.

*Proof that $D'$ satisfies Condition (b); that is, $\max D' \leq n' - 1$:*

Consider first the case that $D$ is flat. Then $\max D' = d$ and $n' = n - 2d + 5$. By Observation 1 $n \geq 3d$. Thus $d = \max D' \leq n' - 5$.

Now consider the case that $D$ is typical. Then $n' = n - k + 2$. Assume for the sake of contradiction that $b \in D'$ and $b \geq n' = n - k + 2$. If $b \notin D$, then $b = \ell - k + 2 \leq n - 1 - k + 2 \leq n - k + 1 < n'$ which is the desired contradiction. Thus we may assume $b \in D$. We will derive a contradiction by demonstrating that in this case Lemma 14 would apply to $D$. Let $b = n - r - 1$. $b \neq n - 1$ since Lemma 13 would apply to $D$, thus $r \geq 1$. Furthermore, since $b \geq n - k + 2$, $r \leq k - 3$. By Observation 2, $n_2 \geq \alpha - 1$. Having $D$ typical then implies $n_2 \geq k - 1$. Thus $1 \leq r < n_2$. Having, $r \leq k - 3$ implies $k - r > 2$ and $\ell - r \geq \ell - k + 3 > 2$ since $\ell > k$. Thus Lemma 14 with $r := r$, $x := \ell$, $y := k$, and $n - r - 1 = b$ applies to $D$, which is the desired contradiction.

Now consider the case that $D$ is special. Then we know that $n' = n - 3$, $k = 4$ and $\ell \geq 5$. Assume for the sake of contradiction that $b \in D'$ and $b \geq n' = n - 3$. If $b \notin D$, then $b = \ell - 2 \geq n - 3$. That however is not possible, since it implies that $\ell = n - 1 \in D$, in which case Lemma 13 would apply to $D$. Thus we may assume $b \in D$, and $b = n - r - 1$, where either $r = 1$ or $r = 2$. Thus $1 \leq r \leq n_2$, and $k - r \geq 2$, since $k = 4$. Similarly, $\ell - r \geq 3$ since $\ell \geq 5$. Thus Lemma 14 with $r := r$, $x := \ell$, $y := k$, and $n - r - 1 = b$ applies to $D$, which is the desired contradiction.

*Proof that $D'$ satisfies Condition (d); that is, $D' \notin \{\langle 2^{\langle n'-4 \rangle}, d^{\langle 4 \rangle} \rangle : d \geq 5\}$:*

If $D$ is flat or special $D$, then $n'_d \geq 1$. Thus if $D' \in \{\langle 2^{\langle n'-4 \rangle}, d, d, d, d \rangle : d \geq 5\}$ then $D$ is typical. However, in that case $D \in \{\langle 2^{\langle n-5 \rangle}, \alpha, d, d, d, d + \alpha - 2 \rangle : \alpha \geq 3\}$ which is not possible since Lemma 11 would apply to $D$.

*Proof that $D'$ satisfies Condition (e); that is, if $D'$ is even, then $n'_2 \geq \frac{n'}{3} + 1$:*

An even degree sequence is *bad* if it satisfies Conditions (a)-(d) but not Condition (e). We need to prove that $D'$ is not bad. We start with the following observation.

**Observation 3.** *If $D'$ does not satisfy Condition (e); that is, if $D'$ is bad, then $n'_4 \geq 4$.*

*Proof.* Since $D'$ satisfies Condition (a), $4n' - 6 \geq 2n'_2 + 4n'_4 + 6(n' - n'_4 - n'_2)$ which simplifies to $n'_2 \geq \frac{n' - n'_4 + 3}{2}$. If $D'$ is bad, then $n'_2 < \frac{n'}{3} + 1$, by definition. Thus $\frac{n' - n'_4 + 3}{2} \leq n'_2 < \frac{n'}{3} + 1$, and consequently $\frac{n' - n'_4 + 3}{2} < \frac{n'}{3} + 1$ giving $n'_4 > \frac{n' + 3}{3}$. Since $D'$ is even and satisfies Conditions (a)-(c), the only such sequences with $n' \leq 5$ elements are $\langle 2, 2, 2 \rangle$ and $\langle 2, 2, 2, 4, 4 \rangle$. However, these sequences are not bad, and thus $n' \geq 6$. Therefore, having $n'_4 > \frac{n' + 3}{3}$ implies that $n'_4 \geq 4$. □

We are now ready to prove that $D'$ is not bad; that is, that $D'$ satisfies Condition (e). To do so, it suffices to demonstrate that $n'_4 \leq 3$, by Observation 3.

Case 1. $D$ is flat: By Observation 1, $d \geq 5$, and thus there is exactly one 4 in $D'$. Therefore $D'$ is not bad.

Case 2. $D$ is typical and even: Then $n'_4 \leq n_4 + 1$. Since $D$ is typical, $n_4 \leq 2$. Thus $n'_4 \leq 3$ and by the Observation 3 $D'$ is not bad.

Case 3. $D$ is typical and not even: If $D$ has at least three odd numbers, or $k(=\alpha)$ is even, or $\ell$ is even, then $D'$ is not even and thus is not bad. If $D'$ is bad, then $n_4 \geq 3$ since $n'_4 \leq n_4 + 1$ and since each bad sequence has at least four 4's. Thus the only remaining case is that $n_4 \geq 3$, $\alpha = 3$ (since $\alpha$ is odd and $4 \in D$), $\ell$ is odd and $D$ has exactly two odd numbers. Then $\ell \neq 3$, since in that situation we would have chosen $k = 3$ and would have changed $\ell$ to 4 (reversing the roles of $\ell$ and $k$). However, in that case $k$ would be even which was ruled out above. Thus $k = x$ where $x$ is odd and $x \geq 5$. However, in that case $D$ would be special and not even.

Case 4. $D$ is special and even: Then $n' = n - 3$ and $n'_2 = n_2 - 2 + 1 = n_2 - 1$. Since $D \in \mathbf{\Delta}$, $n_2 \geq \frac{n}{3} + 1$. Thus $n'_2 + 1 \geq \frac{n'+3}{3} + 1$ which simplifies to $n'_2 \geq \frac{n'}{3} + 1$. Therefore, $D'$ is not bad.

Case 5. $D$ is special and not even: Then $3 \in D'$ and $D'$ cannot be bad.

We have verified that $D'$ satisfies Conditions (a)–(e), thus completing the proof of the claim. □

Together, Lemma 7 and Lemma 15 prove Theorem 1.

## 5 Algorithmics

Theorem 1 provides an easy $O(n)$ time algorithm for recognizing the degree sequences of 2-trees simply by verifying Conditions (a)–(e). When a sequence is realizable as a 2-tree $G$, the proof of Lemma 15 leads to an $O(n)$ time algorithm for constructing $G$ that we sketch here.

First, observe that the elements of $D$ are all integers in $\{2, \ldots, n-1\}$ and can therefore be sorted in $O(n)$ time [6]. We can then represent the sequence $D$ using *run-length encoding*. That is, we use a list of pairs $\{(d_i, r_i) : 1 \leq i \leq p\}$ where $r_i$ is the multiplicity of the element $d_i$ in $D$. We keep this list sorted by the $d_j$ values at all times during the algorithm. (Here $p$ is the number of distinct values in $D$.) Along with this encoding we keep two counters. The counter $n = \sum_{i=1}^{p} r_i$ is the number of elements in the sequence and the counter $n_0$ is the total number of odd values in $D$.

The inductive proof in Lemma 15 results in a recursive algorithm that makes $O(n)$ recursive calls. Each invocation takes as input the sequence $D$ and a pointer to the node in the linked list containing the pair $(d_i, r_i)$ with $d_i = \ell$. An invocation performs four steps:

1. check if the sequence $D$ conforms to any of the base cases in Lemmas 8–14,

2. determine whether $D$ is flat, typical or special,

3. select a value $k \in D$, and

4. remove some number of 2's, 4's and/or $d$'s from $D$ and reduce the values of at most two other elements in $D$ before recursing.

Note that each base case in Lemmas 8–12 has a run-length encodings of $O(1)$ size, and thus can be checked in $O(1)$ time. The base case of Lemma 13 does not necessarily have a run-length encoding of $O(1)$ size, but can be checked in $O(1)$ time by checking if $d_k = n - 1$. Lemma 14 also does not necessarily have a run-length encoding of $O(1)$ size but it can be verified that, if the conditions of Lemma 14 hold then they hold with $n - r - 1$ being selected from among the three largest values in $D$. Thus, the conditions of Lemma 14 can be checked in $O(1)$ time by considering (at most) $d_p$, $d_{p-1}$, $d_{p-2}$, and $r_1 (= n_2)$. Thus, in $O(1)$ time we can check if any of the base cases described in Lemmas 8–14 apply to the sequence $D$. Furthermore, each of the constructions in the base cases are explicit and can easily be accomplished in $O(n)$ time. Whether $D$ is flat can be determined in $O(1)$ time since a flat sequence has a run-length encoding with at most two elements. Whether $D$ is special can be determined by counting the number of 4's in $D$ (which is given by $r_2$ or $r_3$) and by checking if $D$ contains exactly two odd numbers, one of which is 3. This can also be done in $O(1)$ time by checking the values of $d_2$ and $n_o$.

The value $k \in D$ can be selected in $O(1)$ time since a careful inspection of the proof reveals that $k$ is either $d_2$ or $d_3$.

Finally, 2's, 4's and $d$'s can be removed from the sequence in $O(1)$ time by reducing the values of $r_1$, $r_2$ and/or $r_3$ as appropriate. Reducing the values of $k$ and $\ell$ causes these values to move forward in the run-length encoding. (Recall that our list must remain sorted according to the $d_i$ values.) However, this can easily be implemented in $O(k)$ time and causes the sum of the sequence to decrease by $2k - 4$. Since the initial sum of the sequence is $4n - 6$ this means that the total time spent on reducing values during all steps is $O(n)$. Thus, the entire algorithm runs in $O(n)$ time.

## 6 Conclusion

Prior to this work, the degree sequences of $k$-trees were characterized for $k = 1$ only, that is for trees. In this paper, we settle the $k = 2$ case. An obvious direction for future work is to characterize the degree sequences of $k$-trees for $k \geq 3$. We conclude this paper with some arguments highlighting why, in the general case at least, this may be difficult.

A related, and less well-known concept, is that of a degree set. The *degree set* of a graph is the set of the degrees of its vertices. Unlike degree sequences, degree sets contain no information about the multiplicities of the degrees. Kapoor *et al.* [17] characterized the degree sets of $n$-vertex simple graphs, $n$-vertex trees, $n$-vertex outerplanar graphs, and $n$-vertex planar graphs. Degree sets of $k$-trees have been studied by Winkler [23, 24] and Duke and Winkler [7, 8, 9] who prove that all but finitely many degree sets are realizable by $k$-trees..

Characterizing degree sequences is more difficult than characterizing degree sets. A characterization of the degree sets for a class of graphs, can be inferred from a characterization of the degree sequences for that class. For example, Theorem 1 implies immediately that a set of integers $S$ is the degree set of some 2-tree if and only if the minimum element in $S$ is 2, which is a result of Duke and Winkler[4] They also characterize the degree sets of 3-trees and 4-trees as the sets with minimum element 3 and 4 respectively, except for the set $\{4, 7, 8\}$ which is not realizable as a 4-tree. Despite their effort however, no characterization of the degree sets of $k$-trees is known, suggesting that characterizing degree sequences of $k$-trees may be complicated. To appreciate the difficulty, even in the case of degree sets, consider the following result by Duke and Winkler [9]: Let $S := \{k, k+d-1, k+d+r-1\}$ for some positive integers $k \geq d+r$. Then, for $r < d$, $S$ is the degree sequence of some $k$-tree if and only if $d \equiv 1 \pmod{r}$. For $r = d$, $S$ is the degree set of some $k$-tree if and only if $d = 2$. For $r > d$, no set of necessary and sufficient conditions has been found.

The following three well-known conditions are necessary for a sequence of positive integers $D$ to be the degree sequence of a $k$-tree:

(i) $\sum D = 2kn - k(k+1)$,

(ii) $\max D \leq n - 1$,

(iii) $\min D = k$ and $n_k \geq 2$.

Let $D = \langle d_1, \ldots, d_n \rangle$ be a sequence of $n$ positive integers, for some $n \geq k \geq 2$. Let $D'$ be the sequence of $n' := n+1$ integers $\langle d_1+1, ..., d_n+1, n'-1 \rangle$. Then $D'$ is realizable as a $k$-tree if and only if $D$ is realizable as a $(k-1)$-tree. This is because the neighbourhood of every vertex in a $k$-tree induces a $(k-1)$-tree, and adding a new vertex adjacent to every vertex of a $(k-1)$-tree produces a $k$-tree.

Thus if $D$ satisfies conditions (i)-(iii) for $k-1$ but $D$ is not realizable as a $(k-1)$-tree, then $D'$ satisfies (i)-(iii) for $k$ but $D'$ is not realizable as a $k$-tree. For example, by Theorem 1, if a sequence $D = \langle 3^{\langle n_3 \rangle}, a_1, \ldots, a_{n-n_3-1}, n-1 \rangle$ with each $\langle a_1, \ldots, a_{n-n_3-1} \rangle$ odd is realizable as a 3-tree, then $n_3 \geq \frac{n-1}{3} + 1$. In general, the following conditions are necessary for $D$ to be the degree sequence of some $k$-tree:

(iv) if $k \geq 2$, then $D \notin \{\langle k^{\langle n-k-2 \rangle}, (d+k-2)^{\langle 4 \rangle}, (n-1)^{\langle k-2 \rangle} \rangle : d \geq 5\}$,

(v) if $k \geq 2$ and $D = \langle k^{n_k}, a_1, \ldots, a_{n-n_k-k+2}, (n-1)^{\langle k-2 \rangle} \rangle$ with each $a_i \equiv k \pmod{2}$, then $n_k \geq \frac{n-k+2}{3} + 1$.

---
[4]Duke and Winkler use an equivalent definition of $k$-trees in terms of $(k+1)$-uniform hypergraphs. Their results, quoted here, are translated to match the definitions used in this paper.

Moreover, every sequence $\langle k^{n_k}, a_1, \ldots, a_{n-n_k-k+2}, (n-1)^{\langle k-2 \rangle} \rangle$ that meets conditions (i)-(iv) is a degree sequence of a $k$-tree. Finally, Lotker *et al.* [19] contribute another non-trivial necessary condition:

(vi) if $k \geq 2$ and $d := \frac{k(n+1)}{k+2}$ is a positive integer, then $D \neq \langle k^{\langle n-k-2 \rangle}, d^{\langle k+2 \rangle} \rangle$.

## Acknowledgments

This research was initiated at The 21st Bellairs Winter Workshop on Computational Geometry, January 27–February 3, 2006. The authors are grateful to Godfried Toussaint for organizing the workshop and to the other workshop participants for providing a stimulating working environment. Thanks to Yihui Tang and Daming Xu for obtaining and translating the paper by Li and Mao [18].